\documentclass[11pt]{article}
\usepackage{soul}
\usepackage{jheppub}
\usepackage{amsfonts}
\usepackage{amsmath}
\usepackage{amsthm}
\allowdisplaybreaks[4]         
\usepackage{amssymb}   
\usepackage{euscript}   
\usepackage{cleveref}    
\usepackage[dvipsnames]{xcolor}          
\usepackage{tensor}     
\usepackage{graphicx}
\usepackage{caption}

\usepackage{subcaption}   
\usepackage{hyperref}
\usepackage{enumerate}
\newcommand{\req}[1]{(\ref{#1})} %{Eq.\thinspace(\ref{#1})}  

\newcommand{\bea}{\begin{eqnarray}}
\newcommand{\eea}{\end{eqnarray}}
\newcommand{\peff}{p_{\rm eff}}

\newcommand{\ba}{\begin{eqnarray}}
\usepackage{braket}
\newcommand{\ea}{\end{eqnarray}}

\newcommand{\beq}{\begin{equation}}
\newcommand{\eeq}{\end{equation} }
\newcommand{\beqa}{\begin{eqnarray}}

\newcommand{\eeqa}{\end{eqnarray}}
\newcommand{\beqar}{\begin{eqnarray*}}
\newcommand{\eeqar}{\end{eqnarray*}}

\newcommand{\be}{\begin{equation}}
\newcommand{\ee}{\end{equation}}
\renewcommand{\req}[1]{(\ref{#1})}

\newcommand{\eg}{{\it e.g.,}\ }

\definecolor{shadecolor}{rgb}{.25,.25,.25}

\usepackage{caption}
\usepackage{subcaption}

 \title{ \boldmath  Kasner eons in Lovelock black holes}

\author[]{Pablo Bueno,}
\author[]{Pablo A. Cano,}
\author[]{Robie A. Hennigar,}
\author[]{Ming-Da Li}

\affiliation[]{Departament de Física Quàntica i Astrofísica, Institut de Ciències del Cosmos\\ Universitat de Barcelona, Martí i Franquès 1, E-08028 Barcelona, Spain}

\emailAdd{pablobueno@ub.edu}
\emailAdd{pablo.cano@icc.ub.edu}
\emailAdd{robie.hennigar@icc.ub.edu}
\emailAdd{limd23@icc.ub.edu}

\date{\today}
\abstract{In the vicinity of space-like singularities, general relativity predicts that the metric behaves, at each point, as a Kasner space which undergoes a series of ``Kasner epochs'' and ``eras'' characterized by certain transition rules. The period during which this process takes place defines a ``Kasner eon'', which comes to an end when higher-curvature or quantum effects become relevant. %When this happens, the Einsteinian ``Kasner eon'' comes to an end. 
When higher-curvature densities are included in the action, spacetime can undergo transitions into additional Kasner eons. During each eon, the metric behaves locally as a Kasner solution to the higher-curvature density controlling the dynamics. In this paper we identify the presence of Kasner eons in the interior of static and spherically symmetric Lovelock gravity black holes. We determine the conditions under which eons occur and study the Kasner metrics which characterize them, as well as the transitions between them. We show that the null energy condition implies a monotonicity property for the effective Kasner exponent at the end of the Einsteinian eon. We also characterize the Kasner solutions of more general higher-curvature theories of gravity. In particular, we observe that the Einstein gravity condition that the sum of the Kasner exponents adds up to one, $\sum_{i=1}^{D-1}p_i=1$, admits a universal generalization in the form of a family of Kasner metrics satisfying $\sum_{i=1}^{D-1}p_i=2n -1$ which exists for any order-$n$ higher-curvature density and in general dimensions.
}

\begin{document} 
\setlength{\parskip}{0pt}

\maketitle

\flushbottom
%\newpage
%%%%%%%%%%%%%%%%%%%%%%%%%%%%%%

\section{Introduction}

Curvature singularities occur ubiquitously in General Relativity, both in cosmology and in black holes~\cite{Hawking:1973uf}. They represent a complete breakdown of the classical theory and mark the limits of its predictive power. It is commonly thought that quantum gravity will resolve singularities, but very little is known about this in practice. What is certain is that quantum effects will play an essential role near singularities and understanding these effects along with how, if, or what kinds of singularities can be resolved are fundamentally important questions.

A remarkable fact is that the ``death throes'' of General Relativity contain a very universal structure. As shown by Belinski, Khalatnikov and Lifshitz (BKL), General Relativity admits generic spacelike singularities that are \textit{ultralocal} and \textit{oscillatory} characterized by an infinite sequence of Kasner \textit{epochs} and \textit{eras}~\cite{Belinsky:1970ew}. Importantly, the onset of BKL dynamics can occur already at curvature scales where the classical theory should remain reliable. These universal features may provide a path to understand quantum effects on singularities. 

In the approach to a spacelike singularity there is a  decoupling of spatial points, leading to an emergent ultralocality where each spatial point evolves independently from the others. In this regime the Universe is described by a generalized Kasner metric, which is similar to the familiar Kasner solution, 
\be \label{kasn0}
\mathrm{d}s^2=-\mathrm{d} t^2+\sum_{i=1}^{(D-1)}t^{2p_i} \mathrm{d} x_i^2 \, , \quad \sum_{i = 1}^{(D-1)} p_i =  \sum_{i = 1}^{(D-1)} p_i^2 = 1 \, ,
\ee
with the difference that the exponents $p_i$ are permitted to depend on space. The corresponding period of time where the generalized Kasner metric remains a good approximation is known as a Kasner epoch. The ultralocal regime is punctuated by brief transitions driven by spatial curvature wherein the universe transitions from one  Kasner epoch to another. Kasner eras comprise larger time intervals and are made up of several epochs. The defining feature of an era is that the transitions between epochs involve the repeated swapping of the smallest two Kasner exponents, while the remaining exponents monotonically decrease. The sequence of transitions, and the corresponding changes in expansion and contraction of the universe, leads to the oscillatory dynamics in the approach to the singularity.

One manifestation of quantum gravitational effects, common to many approaches, is the appearance of higher-derivative corrections to the Einstein-Hilbert action~\cite{tHooft:1974toh,Goroff:1985th,Gross:1986mw,Grisaru:1986vi,Sakharov:1967nyk,Visser:2002ew,Endlich:2017tqa}. In the approach to a singularity, these terms will ultimately become important and will lead to drastic modifications of the BKL analysis. While there has been growing interesting in understanding aspects of the black hole interior and singularity, e.g.~\cite{Frenkel:2020ysx, Hartnoll:2020rwq, Cai:2020wrp, Caceres:2022smh, DeClerck:2023fax, Arean:2024pzo, Cai:2024ltu}, there as yet have been very few studies concerning the implications of higher-curvature corrections for ultralocality and the chaotic, oscillatory dynamics predicted by BKL. It was recently argued by three of us that the consideration of higher-derivative corrections naturally introduces the concept of an {\it eon}: periods which are dominated by emergent physics at each energy scale~\cite{Bueno:2024fzg}. From this perspective, the period in which the entire BKL dynamics of General Relativity occurs constitutes the \textit{Einsteinian eon}. Different ways by which the Einsteinian eon may come to an end were explored in~\cite{Bueno:2024fzg}, including the possibility of finite volume singularities, inner horizons, or additional eons.

It was proposed in~\cite{Bueno:2024fzg} that under certain circumstances, such as a hierarchy of energy scales, additional eons could appear. During the additional eons, one could imagine modified BKL-like dynamics, consisting of epochs and eras, but with the Kasner exponents obeying modified constraints and transition rules dictated by the modified gravitational equations. Exploring this idea concretely is a rather difficult but interesting problem requiring the extenstion of the BKL analysis to higher-curvature theories of gravity. As evidence for this idea, a toy model was explored consisting of the interior of a spherically symmetric black hole in Gauss-Bonnet gravity. The Gauss-Bonnet theory introduces a new energy scale by its coupling constant $\lambda$. For scales $M \gg r \gg \sqrt{\lambda}$ it was observed that the interior geometry is given to a good approximation by a Kasner solution of Einstein gravity. However, for $r \ll \sqrt{\lambda}$ a transition occurs and the geometry is then given by a Kasner solution of the Gauss-Bonnet theory --- this is a Gauss-Bonnet eon consisting of a single Kasner epoch. The interior solution provides a smooth connection between the Einsteinian and Gauss-Bonnet eons. 

The purpose of this paper is to further explore the ideas of~\cite{Bueno:2024fzg} as a step toward a more complete understanding of how higher-curvature corrections alter and supplement the results of the BKL analysis. We begin in section~\ref{kas_class} by performing a classification of Kasner solutions of various higher-curvature theories. We point out an apparently universal feature, namely, that for every density involving $n$ powers of the Riemann tensor, there exists a family of Kasner solutions for which the sum of the Kasner exponents equals $2n-1$, generalizing the Einstein gravity result. We then, in section~\ref{Love_eons}, focus on Lovelock theory where analytical black hole solutions are available and study the Kasner geometries that emerge in the black hole interior. Introducing an effective Kasner exponent which is constant during periods where the metric is approximately Kasner, we study the existence of eons in these black holes illustrating how the Einsteinian eon can be followed either by additional Lovelock eons or terminate in a finite volume singularity.\footnote{Therefore, in this paper we do not focus on the cases in which the higher-curvature terms give rise to additional inner horizons. In that context, it has been recently shown that adding infinite towers of higher-curvature corrections can lead to a full resolution of the  Schwarzschild black hole singularity in $D\geq 5$ \cite{Bueno:2024dgm,Konoplya:2024hfg,DiFilippo:2024mwm,Konoplya:2024kih,Ma:2024olw}. }  Finally, in section~\ref{end_eon} we make some more general remarks concerning the end of an eon. We derive perturbative formulas governing the behaviour of the effective Kasner exponent at the end of an eon, which connects our results with the more traditional effective field theory program. Finally, we show that if the effective stress tensor generated by the higher-curvature terms respects the null energy condition then the effective Kasner exponent exhibits a monotonic behaviour at the end of the Einsteinian eon.  In appendix \ref{GB}, we perform a detailed analysis of the different types of interiors which arise as a function of the sign and magnitude of the gravitational couplings for Gauss-Bonnet gravity in general dimensions as well as for cubic Lovelock gravity in $D=7$.

\section{Kasner solutions in higher-curvature gravity}
\label{kas_class}
We are interested in Kasner solutions to higher-curvature theories of gravity.\footnote{For other examples of investigations of Kasner solutions for specific higher-curvature theories see, e.g.,~\cite{Deruelle:1989he, Clifton:2006kc, Middleton:2010bv}.} The Kasner metric is given by \req{kasn0}, and the equations of motion of a given theory constrain the ``Kasner exponents'' $p_i$ in different ways. %In order to find solutions of this type, one can either insert an ansatz of the above form in the corresponding equations of motion and find the conditions for the Kasner exponents. Alternatively, we can consider an ansatz with $D$ arbitrary functions of the form
%\begin{equation}\label{gK}
%\mathrm{d}s^2=-N(t )\mathrm{d}t^2+ \sum_{i=1}^{(D-1)} a_i(t )\mathrm{d}x_i^2\, ,
%\end{equation}
%find the on-shell action 
%\begin{equation}
%S[N, a_i] \equiv  \int \mathrm{d}t L[N,a_i] \, , \quad \text{where} \quad  L[N,a_i]\equiv \sqrt{N(t) a_1(t)\cdots a_{D-1}(t)} \left. \mathcal{L}\right|_{\req{gK}}\, ,
%\end{equation}
%and vary it with respect to those functions. The result reads
%As a consequence of the invariance of the ansatz under relabellings $i \leftrightarrow j$ with $i \neq j$, the full equations of motion of a given theory will always reduce to a pair of independent equations for $N(t)$ and one of the scale factors, respectively. Varying the effective action $S[N,a_i]$ with respect to $N(t)$ and $a_i(t)$ will yield such equations. Indeed, one finds %\comment{check factors}
%\begin{equation}
%\frac{N^2}{\sqrt{N a_1\cdots a_{D-1}}}\frac{\delta S[N,a_i]}{\delta N}=\mathcal{E}_{tt} [N,a_i] \, , \quad -\frac{a_i^2}{\sqrt{N a_1\cdots a_{D-1}}}\frac{\delta S[N,a_i]}{\delta a_i}=\mathcal{E}_{ii}[N,a_i] \, , 
%\end{equation}
%where $\mathcal{E}_{ab}[N,a_i]\equiv \left. \frac{1}{\sqrt{|g|}}\frac{\delta S}{\delta g^{ab}} \right|_{\req{gK}}$ are the field equations of the theory evaluated on the ansatz \req{gK}.  Hence, solving the Euler-Lagrange equations of the effective Lagrangian associated to $N(t)$ and $a_i(t)$ is equivalent to solving the full non-linear equations of motion. Once, we have the equations, we can set 
%$
%N(t)=1$,  $a_i(t)=t^{2p_i}\, ,
%$
%and solve them for $p_i$. 
In order to characterize such constraints, it is convenient to introduce the parameters $\mu,\nu$, as
\begin{equation}\label{munu}
\mu \equiv \sum_{i=1}^{(D-1)} p_i\, , \quad \nu\equiv  \sum_{i=1}^{(D-1)} p_i^2\, ,
\end{equation}
%which turn out to be useful in order to characterize the solutions. 
These are invariant under arbitrary permutations of pairs of Kasner exponents. In the case of Einstein gravity,
\begin{equation}
I=\int \mathrm{d}^Dx \sqrt{|g|} R\, ,
\end{equation}
there exists a $(D-3)$-parametric family of solutions determined by the conditions
%, after imposing \req{Nai}, that $\mathcal{E}_{tt}$ enforces the condition $ \nu=\mu^2$, whereas $\mathcal{E}_{11}$ imposes $\mu =1$. Therefore, the most general Kasner solution of Einstein gravity is characterized by any set of exponents satisfying the conditions
\begin{equation}\label{KasnerEinstein}
 \mu=1 \, , \quad \nu=1\, .
\end{equation}
Additionally, there exists an isolated solution corresponding to 
$
p_1=p_2=\dots=p_{(D-1)}=0\, ,
$
which is nothing but $D$-dimensional Minkowski spacetime.

%The symmetry of the equations of motion under permutations of pairs of Kasner exponents makes it clear that the conditions should always be expresable in terms of invariants such as $\mu$, $\nu$ and similar combinations such as $\sum_{i \neq j}^{D-1}p_i^{2} p_j$, $\sum_{i \neq j \neq k}^{D-1}p_i p_jp_k$, etc.

%\subsection{On-shell action method for Kasner solutions}
%In order to characterize the Kasner solutions of a given higher-curvature density, it is convenient to use an on-shell action method which 
%\comment{symmetric criticality blah blah}. Let us explain it here.
We are interested in Kasner solutions of the form \req{kasn0}.
%\begin{equation}\label{kasn}
%\mathrm{d}s^2=-\mathrm{d} t^2+\sum_{i=1}^{(D-1)}t^{2p_i} \mathrm{d} x_i^2 \, .
%\end{equation}
%The equations of motion of a given theory constrain the ``Kasner exponents'' $p_i$ in different ways. I
In order to find solutions of this type, one can either insert an ansatz of the above form in the corresponding equations of motion and find the conditions for the Kasner exponents. Alternatively, we can consider an ansatz with $D$ arbitrary functions of the form
\begin{equation}\label{gK}
\mathrm{d}s^2=-N(t )\mathrm{d}t^2+ \sum_{i=1}^{(D-1)} a_i(t )\mathrm{d}x_i^2\, ,
\end{equation}
find the on-shell action 
\begin{equation}
S[N, a_i] \equiv  \int \mathrm{d}t L[N,a_i] \, , \quad \text{where} \quad  L[N,a_i]\equiv \sqrt{N(t) a_1(t)\cdots a_{D-1}(t)} \left. \mathcal{L}\right|_{\req{gK}}\, ,
\end{equation}
and vary it with respect to those functions. The result reads
%As a consequence of the invariance of the ansatz under relabellings $i \leftrightarrow j$ with $i \neq j$, the full equations of motion of a given theory will always reduce to a pair of independent equations for $N(t)$ and one of the scale factors, respectively. Varying the effective action $S[N,a_i]$ with respect to $N(t)$ and $a_i(t)$ will yield such equations. Indeed, one finds %\comment{check factors}
\begin{equation}
\frac{N^2}{\sqrt{N a_1\cdots a_{D-1}}}\frac{\delta S[N,a_i]}{\delta N}=\mathcal{E}_{tt} [N,a_i] \, , \quad -\frac{a_i^2}{\sqrt{N a_1\cdots a_{D-1}}}\frac{\delta S[N,a_i]}{\delta a_i}=\mathcal{E}_{ii}[N,a_i] \, , 
\end{equation}
where $\mathcal{E}_{ab}[N,a_i]\equiv \left. \frac{1}{\sqrt{|g|}}\frac{\delta S}{\delta g^{ab}} \right|_{\req{gK}}$ are the field equations of the theory evaluated on the ansatz \req{gK}.  Hence, solving the Euler-Lagrange equations of the effective Lagrangian associated to $N(t)$ and $a_i(t)$ is equivalent to solving the full non-linear equations of motion --- see \eg \cite{Deser:2003up,Palais:1979rca,Bueno:2017qce,Arciniega:2018tnn,Bueno:2019ycr} for previous instances in which similar methods were used for finding solutions with different isometries. Once, we have the equations, we can set 
$
N(t)=1$,  $a_i(t)=t^{2p_i}\, ,
$
and solve them for $p_i$. % In the following subsections we use this method to obtain various new families of Kasner solutions .

\subsection{A universal feature}
In the following subsections we use the above method to characterize the Kasner metrics of various higher-curvature theories in the absence of matter. Our list if not fully exhaustive as, in certain cases, there exist  isolated sets of solutions which cannot be easily characterized in general dimensions and for arbitrary curvature orders. Additionally, for a given curvature order one can either study the Kasner solutions for general values of the  coupling constants or, alternatively, study the solutions of isolated densities. The first approach gives rise to very messy expressions as soon as we move beyond quadratic curvature order. Just like for Einstein gravity, in each case we find the existence of broad families of solutions, corresponding to hypersurfaces in the $\{p_i \}_{i=1,\dots, D-1}$ hyperplane characterized by certain constraints on the values of $\mu$ and $\nu$, as well as sets of isolated solutions which correspond to points in such hyperplane. 

Our analysis reveals an interesting general feature.  Namely, we observe that the family of metrics characterized by the Einsteinian conditions \req{KasnerEinstein}
gets generalized, for general order-$n$ densities, 
\begin{equation}
I=\int \mathrm{d}^Dx \sqrt{|g|} \text{Riem}^n\, ,
\end{equation}
to a family of solutions characterized by a condition of the form
%are solutions to general higher-curvature theories. In addition to this, we find that, for any linear combination of densities involving order-$n$ densities (namely, built from general contractions of $n$ Riemann tensors),  there always exist an additional class of solutions fulfilling the conditions
\begin{equation}\label{Kasnerg}
 \mu=2n -1 \, , 
\end{equation}
plus an additional, more complicated, constraint which can be written in the form
\begin{equation}\label{nuni}
\nu=\nu(\{\alpha_i\},p_3,p_4,\dots,p_{(D-1)})\, ,
\end{equation}
where $\nu$ is in general a complicated function of the relative gravitational couplings $\alpha_i$ and $(D-3)$ of the Kasner exponents. Therefore, we find that $\mu$ does not depend on the spacetime dimension and its dependence on the order of the density is a remarkably simple generalization of the Einstein gravity case, corresponding to $\mu=1$. We have verified this feature for $f(R)$, quadratic, cubic and Lovelock gravities in various dimensions which makes us confident that this indeed a universal property of higher-curvature densities. It is then more than tempting to conjecture that assuming some generalized BKL-type behavior persists in the interior of generic black holes dominated by higher-curvature interactions, the corresponding Kasner exponents characterizing the spacetime metric at each point will satisfy \req{Kasnerg} and \req{nuni} instead of the usual conditions \req{KasnerEinstein}.

%In addition to this class, we find additional solutions corresponding to isolated points in the $\{p_i \}_{i=1,\dots, D-1}$ hyperplane. These are often (but not always) characterized by a partial or full degeneracy of all the Kasner exponents,
%$
%p_1=p_2=\dots=p_{(D-1)}\, .
%$
\subsection{Explicit examples}
\subsubsection{$f(R)$ gravity}
Consider a density consisting of an arbitrary power of the Ricci scalar, namely
\begin{equation}
I=\int \mathrm{d}^Dx \sqrt{|g|} R^n\, .
\end{equation}
Interestingly, whenever $n \geq 2$, this theory admits Kasner solutions whose exponents satisfy a single relation (instead of two), namely,
\begin{equation}\label{KasnerfR}
\nu=\mu(2- \mu)\, , 
\end{equation}
where $\mu$ can in principle take any real value, but it is constrained to the range $0\leq \mu \leq 2$ in order for the metric to remain real-valued. In $D$ dimensions, this represents a $(D-2)$-parametric family of solutions. This obviously includes the Einstein gravity set \req{KasnerEinstein} as well as Minkowski as particular cases. In addition to this family, there exist ``isolated'' solutions corresponding to
\begin{equation}
p_1=p_2=\dots=p_{(D-1)}=-\frac{(2n-1)(n-1)}{\left(n-\frac{D}{2}\right)}\, ,
\end{equation}
for every $D$ and $n\neq D/2$.

\subsubsection{Quadratic gravities}
The next natural case corresponds to a general quadratic gravity of the form
\begin{equation}\label{quadi}
I=\int \mathrm{d}^Dx \sqrt{|g|} \left[ \alpha_1 R^2 + \alpha_2 R_{ab}R^{ab}+\alpha_3 R_{abcd}R^{abcd}\right]\, .
\end{equation}
%Consider first the case of individual densities. If we set $ \alpha_2= \alpha_3=0$, we are left with a pure $R^2$ term which was studied in the previous subsection. On the other hand, setting $ \alpha_1= \alpha_3=0$,
This theory admits %of the same class as Einstein gravity, namely, satisfying \req{KasnerEinstein}. Additionally, 
a new $(D-3)$-parametric family of solutions satisfying  $\mu=3$ in general dimensions. In $D=4$ this satisfies
\begin{align}\label{4dquad}
\mu=3\, , \quad
\nu=\frac{-3\alpha_1+\alpha_2+7\alpha_3\pm 2\sqrt{2}\sqrt{-(\alpha_2+4\alpha_3)(3\alpha_1+\alpha_2+\alpha_3)}}{\alpha_1+\alpha_2+3\alpha_3}
\, .
\end{align}
In order for the solutions to exist, $\nu$ must be real and positive, which imposes the conditions
\begin{equation}
\{ \alpha_2<-4\alpha_3\, , \, -\frac{1}{3}(\alpha_2+\alpha_3)\leq \alpha_1 < -(\alpha_2+3\alpha_3)\}\, ,
\end{equation}
or, alternatively,
\begin{equation}
\{\alpha_2>-4\alpha_3\, , \, -(\alpha_2+3\alpha_3)< \alpha_1 \leq -\frac{1}{3}(\alpha_2+\alpha_3)\}\, .
\end{equation}
As a consequence, setting any pair of couplings to zero gives rise to invalid solutions. For instance, if we choose $\alpha_2=-4\alpha_3$, which would put the action in the form of a linear combination of $R^2$ with the Gauss-Bonnet density, one would get $\nu=-3$, which is not allowed. Also, setting $\alpha_1=\alpha_3=0$, which would be a pure $R_{ab}R^{ab}$ theory, would yield $\nu=1+2{\rm i } \sqrt{2}$, and $\alpha_1=\alpha_2=0$ would give $\nu=\frac{1}{3}[ 7+4{\rm i } \sqrt{2}]$ which is not valid either. On the other hand, the combination $\alpha_2=-3\alpha_1-\alpha_3$, which corresponds to  a linear combination of a Weyl tensor-squared term plus a $R_{abcd}R^{abcd}-R_{ab}R^{ab}$ one, does produce a valid result, namely, $\nu=3$.  In the $D=4$ case, the general quadratic theory also admits a family of solutions  of the same type as Einstein gravity, namely, satisfying \req{KasnerEinstein}.
In addition, there is an isolated solution corresponding to 
\begin{equation}
p_1=p_2=p_3=\frac{1}{2}\, .
\end{equation}

 In higher dimensions, the expression for $\nu$ gets increasingly complicated. For instance, the $D=5$ version of \req{4dquad}  reads
\begin{eqnarray}\label{5dquad}
\mu=3\, , \quad
\nu=\frac{-3 \alpha_1+\alpha_2+7 \alpha_3-2 \alpha_3 p_3
  p_4\pm \frac{1}{2} \sqrt{A(\alpha_1,\alpha_2,\alpha_3,p_3,p_4)}}{\alpha_1+\alpha_2+3 \alpha_3} ,
\end{eqnarray}
where
\begin{align}
&A(\alpha_1,\alpha_2,\alpha_3,p_3,p_4)\equiv [6 \alpha_1-2 (\alpha_2+\alpha_3 (7-2 p_3 p_4))]^2-4 (\alpha_1+\alpha_2+3 \alpha_3)  \times \\ \notag &\quad\left[9
   \alpha_1+9 \alpha_2+\alpha_3 \left(27-4 p_3 p_4\left(2 \left(p_3
   p_4+(p_3-3) p_3+p_4^2\right)-6 p_4+9\right)\right)\right]\, ,
\end{align}
and where we chose to write $p_1$ and $p_2$ in terms of $\mu,\nu$. For $D\geq 5$, the Einstein gravity family \req{KasnerEinstein} is no longer a solution.  On the other hand, there exist additional isolated solutions satisfying $p_1=p_2=\dots =p_{(D-1)}$ for certain combinations of $\alpha_1,\alpha_2,\alpha_3$ and

%which is the same as condition (10) (except for the different number of kasner parameters) when
%\begin{eqnarray}
%	p_3=\frac{1}{2}(3-p_4\pm\sqrt{-15+6p_4-3p_4^2+\frac{8\alpha_2+32\alpha_3-4\sqrt{2}\sqrt{-((3\alpha_1+\alpha_2+\alpha_3)(\alpha_2+4\alpha_3))}}{\alpha_1+\alpha_2+3\alpha_3}})
%\end{eqnarray}

\subsubsection{Cubic gravities}
The most general cubic Lagrangian contains eight independent densities built from contractions of the Riemann tensor and the metric --- see \eg \cite{Bueno:2016ypa}. In this case, the expressions are rather messy already in $D=4$, and not particularly illuminating. However, we find that in all cases there exists a family of solutions  characterized by the conditions
\begin{equation}
\mu=5\, , \quad \nu=\nu(\{\alpha_i \}, p_3,p_4,\dots, p_{(D-1)})\, ,
\end{equation}
again in agreement with our general observation.

\subsubsection{Lovelock gravities}
Consider now the case of Lagrangians consisting of a Lovelock density of curvature order $n$.
The action is given by 
\begin{equation}\label{LoveF}
I=\int_{\mathcal{M}}\mathrm{d}^{D}x\sqrt{|g|}\mathcal{X}_{2n}\, ,
\end{equation}
where
the dimensionally-extended Euler densities $\mathcal{X}_{2n}$ are given by\footnote{The generalized Kronecker symbol is defined as $\delta^{\mu_1\mu_2\dots \mu_r}_{\nu_1\nu_2\dots\nu_r}\equiv  r!\delta^{[\mu_1}_{\nu_1}\delta^{\mu_2}_{\nu_2}\dots \delta^{\mu_r]}_{\nu_r}$. }
\begin{equation}\label{Euler}
\mathcal{X}_{2n}= \frac{1}{2^{n}}\delta^{\mu_1\dots \mu_{2n}}_{\nu_1\dots \nu_{2n}}R^{\nu_1\nu_2}_{\mu_1\mu_2}\dots R^{\nu_{2n-1}\nu_{2n}}_{\mu_{2n-1}\mu_{2n}}\, .
\end{equation}
The simplest instance beyond the Einstein-Hilbert term corresponds to the Gauss-Bonnet density, which reads
\begin{equation}
\mathcal{X}_{2}= R^2 -4 R_{ab} R^{ab}+R_{abcd}R^{abcd}\, .
\end{equation}
This term contributes non-trivially to the equations of motion for $D\geq 5$. In particular, for $D=5$ we find a family of Kasner solutions characterized by the condition
\begin{equation}
\mu=3\, , \quad p_1=0\, ,
\end{equation}
where one of the Kasner exponents vanishes and the others are free provided the first condition holds. Moving on to $D=6$, we find the families 
\begin{align}
&\mu=3\, , \quad p_1=p_2=0\, , \\
&\mu=3\, , \quad \sum_{i=1}^{5}\frac{1}{p_i}=0\, . 
\end{align}
All of these families were previously identified in \cite{Camanho:2015yqa}, where an exhaustive classification of the Kasner solutions of Lovelock densities in the particular cases of curvature orders satisfying $D=2n+1$ and $D=2n+2$ was performed. Moving on to $D=7$, we find a family of solutions characterized by
\begin{equation}\label{p3p4p5}
\mu=3\, , \quad \nu=\nu(p_3,p_4,p_5)\, ,
\end{equation}
where
\begin{align}\notag
\nu(p_3,p_4,p_5) \equiv & \Big[2 p_3^3 (p_4+p_5+p_6)+2 p_3^2 (p_4+p_5+p_6-3) (p_4+p_5+p_6)+p_3 (p_4+p_5+p_6)\cdot \\ \notag
&   \left(2 \left(p_4^2+p_4(p_5+p_6-3)+p_5^2+p_5 p_6+p_6^2\right)-6 p_5-6 p_6+9\right) \\ \notag &+2 p_4^3
   (p_5+p_6)+2 p_4^2 (p_5+p_6-3) (p_5+p_6)+p_4(p_5+p_6)\cdot \\ \notag & \left(2 \left(p_5 p_6+(p_5-3)
   p_5+p_6^2\right)-6 p_6+9\right)+p_5 p_6\left(2 \left(p_5 p_6+(p_5-3) p_5+p_6^2\right)-6 
   p_6+9\right)\Big] \cdot \\ \notag & \big[p_3 (p_4+p_5+p_6)+p_4 (p_5+p_6)+p_5p_6 \big]^{-1} \, .
\end{align}
Interestingly, all the above solutions reduce to the class
\begin{equation}\label{munuLo}
\mu=3\, , \quad \nu=1+\frac{8}{(D-1)}\, , 
\end{equation}
when $p_2=\dots =p_{D-1}=4/(D-1)$, a case which will be relevant in the analysis of spherically symmetric black hole interiors.

Moving to the case of the cubic Lovelock density,
%\begin{equation}
%\mathcal{X}_{3}= ....\, ,
%\end{equation}
we find for $D=7$ the family of solutions
\begin{equation}
\mu=5\, , \quad p_1=0\, ,
\end{equation}
whereas for $D=8$,
\begin{align}
&\mu=5\, , \quad p_1=p_2=0\, , \\
&\mu=5\, , \quad \sum_{i=1}^{7}\frac{1}{p_i}=0\, .
\end{align}
In both cases, these had previously identified in  \cite{Camanho:2015yqa}. In $D=9$, one finds a family of solutions analogous to \req{p3p4p5} but with 
\begin{equation}
\mu=5\, , \quad \nu=\nu(p_3,p_4,p_5,p_6,p_7)
\end{equation}
and where $\nu(p_3,p_4,p_5,p_6,p_7)$ is not a very illuminating function. Again, in all cases the solutions reduce to a class characterized by
\begin{equation}\label{munuL}
\mu=5\, , \quad \nu=1+\frac{24}{(D-1)}\, , 
\end{equation}
when $p_2=\dots =p_{D-1}=6/(D-1)$.

Both \req{munuLo} and \req{munuL} are particular instances of a more general class of solutions to a general Lovelock density $\mathcal{X}_{2n}$ and in general dimensions, in the particular case in which all exponents but one are equal. This corresponds to
\begin{equation}\label{loveK}
\mu = 2n -1 \, , \quad \nu= 1+\frac{4n(n-1)}{(D-1)}\, , 
\end{equation}
where
\begin{equation}\label{loveK2}
p_1=-\frac{(D-2n-1)}{(D-1)}\, , \quad p_2=\dots =p_{D-1}=\frac{2n}{(D-1)}\, .
\end{equation}
We will see in the following section that this family of solutions arises approximately during certain periods as the singularity of static and spherically symmetric Lovelock black holes is approached.
%\frac{1}{p_3 (p_4+p_5+p_6)+p_4 (p_5+p_6)+p_5p_6}

%where their value depends on the particular theory. Let us see these features in more detail in particular examples.

%n the Einsteinian case, the BKL analysis reveals that it is the first type of solutions the one that controls the near-singularity dynamics. This suggests that it is the first class of solutions the one that controls the near-singularity dynamics

%\subsection{Einstein gravity}
% Let us start with the well-studied case of Einstein gravity, whose Lagrangian reads
%%\begin{equation}
%S=\int \mathrm{d}^Dx \sqrt{|g|} R\, .
%\end{equation}

%In $D$ dimensions, this corresponds to a $(D-3)$-parametric family of solutions.

%\begin{align}
%\mathcal{E}_{tt}&=\frac{1}{2t^2} (\mu^2- \nu) \, , 
%\end{align}
%which imposes $ \nu=\mu^2$. After taking this into account, the other equation reads
%\begin{equation}
%\mathcal{E}_{11}=-\frac{1}{t^{2-2p_1}} (\mu-1) ()
%\end{equation}

%\subsection{Two independent exponents}

\section{Kasner eons from black hole interiors}
\label{Love_eons}
In this section we consider static and spherically symmetric black hole solutions of Lovelock gravity. As the singularity is approached, those spacetimes  undergo one or several Kasner eons through which they locally behave like Kasner solutions of the corresponding higher-curvature density. In the first subsection we define an effective Kasner exponent which becomes constant during an eon for a general static and spherically symmetric spacetime. Then, we use this notion to characterize the presence of Kasner eons in the interior of Lovelock gravity black holes. We determine the conditions under which, depending on the sign and magnitude of the gravitational couplings, the Einsteinian eon is followed by additional higher-curvature eons or terminates in a finite-volume singularity.

\subsection{Effective Kasner exponents}
In this section we focus in the case in which all but one of the exponents coincide with each other, namely, when
\begin{equation}\label{2ind}
p_1 \neq p_2=p_3=\dots =p_{D-1}\, .
\end{equation}
Very often, the metric which describes the near-singularity region of static black hole solutions of higher-curvature theories takes the form \req{kasn0} with Kasner exponents satisfying this  condition. Indeed, consider a general static and spherically symmetric black hole with a single horizon, a spacelike curvature singularity at $r=0$ and with an interior metric described in Schwarzschild coordinates as
\begin{align} \label{bhfN}
\mathrm{d} s^2 =&\, \frac{\mathrm{d}r^2}{f(r)} -N(r) f(r) \mathrm{d} z^2 + r^2 \mathrm{d} \Omega_{(D-2)}^2 \, ,
\end{align}
where $ \mathrm{d} \Omega_{(D-2)}^2$ is the metric of the $(D-2)$-dimensional sphere, and where the two functions $f(r)$ and $N(r)$ behave as
\begin{equation}
f(r) \overset{r\rightarrow 0}{\sim} \, - r^{-s} \, ,\quad N(r) \overset{r\rightarrow 0}{\sim }\, r^{-w}\, ,
\end{equation}
near the singularity, which lies in the future of any infalling observer.  Changing coordinates 
\be 
\mathrm{d} \tau \equiv \frac{\mathrm{d} r}{\sqrt{-f}} \quad \Rightarrow \quad \tau \sim r^{(s+2)/2} \, ,
\ee
the metric becomes
\be \label{taup}
\mathrm{d}s^2 = - \mathrm{d}\tau^2 + \tau^{2p_1} \mathrm{d} z^2 + \tau^{2p_2} \mathrm{d} \Omega_{(D-2)}^2 \, ,
\ee
where the two independent exponents read
\be \label{p1p2}
p_1 = - \frac{(s+w)}{(s+2)} \, , \quad p_2 = \frac{2}{(s+2)} \, .
\ee
Hence, in the vicinity of any point of the $(D-2)$-sphere, the metric takes the usual Kasner form\footnote{Strictly speaking, the interior belongs to the class of Kantowski-Sachs cosmological models, which have $\mathbb{R} \times \mathbb{S}^2$ spatial sections. However, locally, in the vicinity of any point on the two-sphere, the metric can be brought into the usual Kasner form.}
\be \label{taup}
\mathrm{d}s^2 = - \mathrm{d}\tau^2 + \tau^{2p_1} \mathrm{d} z^2 +\sum_{i=2}^{(D-1)}\tau^{2p_i} \mathrm{d}x_i^2 \, ,
\ee
where \req{p1p2} holds and 
\begin{equation}
p_2=p_3=\cdots=p_{D-1}\, . %=p_1+\left(1+\frac{w}{(s+2)}\right)\, .
\end{equation}
In such a general situation, the sums of the Kasner exponents and their squares, as defined in \req{munu}, become
\begin{equation}
\mu= \frac{2(D-2)-(s+w)}{(s+2)}\, , \quad \nu= \frac{4(D-2)+(s+w)^2}{(s+2)^2}\, .
\end{equation}
We can introduce an `effective' Kasner exponent $p_{\rm eff}$ for the $\mathrm{d}z^2$ component of the metric,
\begin{align}\label{effp}
\peff(r) \equiv \frac{r [f(r)N(r)]'}{[2 f(r) - r f'(r)] N(r)} \, ,
\end{align}
so that any time that $f(r) \sim r^{-s}, N(r)\sim r^{-w}$, $\peff(r)$ becomes constant, the metric is locally Kasner and
\be
\label{peffs0} 
p_1 = \peff \, , \quad p_2 = \cdots =  p_{D-1} = \peff + 1+\frac{w}{(s+2)}\, .
\ee

In the present context, Kasner eons correspond to periods during which the interior of black holes in the vicinity of spacelike singularities behave as locally Kasner metrics with approximately constant Kasner exponents satisfying \req{peffs0}. As we will see below, for solutions involving several parametrically distinct length scales, as the singularity is approached, the solutions will transit through various eons characterized by different exponents.

The above expressions get considerably simplified for black holes characterized by a single metric function, namely, those for which $N(r)=1$. In that case, whenever $f(r) \sim r^{-s}$, eons are characterized by Kasner exponents satisfying the conditions
\be
\label{peffs} 
p_1 = \peff=-\frac{s}{(s+2)} \, , \quad p_2 = \cdots =  p_{D-1} = \peff + 1=\frac{2}{(s+2)}\, .
\ee
Consider for instance the case of the $D$-dimensional Schwarzschild black hole, whose metric function reads
\be 
f(r) = 1 - \frac{r_0^{D-3}}{r^{D-3}} \, ,
\ee
where $r_0$ is an integration constant related to the mass of the solution $M$ via
\begin{eqnarray}
r_0^{D-3}&=&\frac{16\pi G M}{(D-2)\Omega_{D-2}}\, ,
\end{eqnarray}
where $\Omega_{D-2}$ is the (dimensionless) volume of the transverse $(D-2)$-sphere. 
In this case, the effective Kasner exponent reads
\be 
\peff(r) = - \frac{D-3}{(D-1) - 2 u^{D-3}} \, , \quad u \equiv r/r_0 \, .
\ee
For $u \ll 1$ --- namely, near the singularity --- this approaches a constant, 
\begin{equation}
\peff = - \frac{(D-3)}{(D-1)}\, ,
\end{equation}
and the metric locally behaves like a Kasner spacetime with exponents 
\begin{equation}\label{EinKas}
p_1= - \frac{(D-3)}{(D-1)}\, , \quad p_2 = \cdots =  p_{D-1} = \frac{2}{(D-1)}\, ,
\end{equation}
which satisfy $\mu=\nu=1$, as expected for Einstein gravity.

%which when combined with~\eqref{peffs} gives the expected result for $D$-dimensional Einstein gravity.

%
%It is well-known that the deep interior of the Schwarzschild black hole is locally a Kasner metric.\footnote{Strictly speaking, the interior belongs to the class of Kantowski-Sachs cosmological models, which have $\mathbb{R} \times \mathbb{S}^2$ spatial sections. However, locally, in the vicinity of any point on the two-sphere, the metric can be brought into the usual Kasner form.} To see this more directly, start from the Schwarzschild metric written in the standard chart
%\begin{align} 
%ds^2 =&\, \frac{dr^2}{f} -f dz^2 + r^2 d \Omega^2 \, ,
%\\
%f =& \, 1 - \frac{r_0}{r} \, .
%\end{align}
%Near the singularity $r < r_0$ so $f < 0$. The `radial' coordinate functions as a time coordinate and we can convert it to proper time. Let $f \sim - r^{-1}$ then 
%\be 
%d \tau \equiv \frac{dr}{\sqrt{-f}} \quad \Rightarrow \quad \tau \sim r^{3/2} \, .
%\ee
%In terms of the proper time, and expanding near the pole of the sphere, the metric reads
%\be 
%ds^2 = - d\tau^2 + \tau^{2p_1} dt^2 + \tau^{2p_2} dx^2 + \tau^{2p_3} dy^2 \, .
%\ee
%which is the standard Kasner form with the exponents
%\be 
%p_1 = - \frac{1}{3} \, , \quad p_2 = p_3 = \frac{2}{3} \, ,
%\ee
%that satisfy
%\be 
%p_1 + p_2 + p_3 = p_1^2 + p_2^2 + p_3^2 = 1\, .
%\ee
%
%More generally, any time the metric behaves like $f \sim r^{-s}$ 

%\rah{More words}
\subsection{Lovelock gravity black holes}
Let us consider now the case of a general Lovelock gravity in $D$ dimensions.
%{\bf Singularities in Lovelock Gravity:}
%Let us start by describing the theory and solutions of interest.
The action is given by 
\begin{equation}\label{LoveF}
I=\int_{\mathcal{M}}\mathrm{d}^{D}x\sqrt{|g|}\mathcal{L}_{\rm Lovelock}\, ,
\end{equation}
where
\begin{align}
{\cal L}_{\rm Lovelock}\equiv \frac{1}{16\pi G} \bigg[ %\frac{(D-1)(D-2)}{L^2}+
R 
+ \sum_{n=2}^{\lfloor (D-1)/2\rfloor}\lambda_n \frac{(D-2n-1)!}{(D-3)!}
%(-1)^{n}
\mathcal{X}_{2n} \bigg]\, ,
\end{align}
is the Lovelock Lagrangian~\cite{Lovelock:1971yv, Lovelock:1972vz}, and where we set the cosmological constant to zero. The dimensionally-extended Euler densities $\mathcal{X}_{2n}$ are defined in \req{Euler}  %\footnote{The generalized Kronecker symbol is defined as $\delta^{\mu_1\mu_2\dots \mu_r}_{\nu_1\nu_2\dots\nu_r}\equiv  r!\delta^{[\mu_1}_{\nu_1}\delta^{\mu_2}_{\nu_2}\dots \delta^{\mu_r]}_{\nu_r}$. }
%\begin{equation}
%\mathcal{X}_{2n}= \frac{1}{2^{n}}\delta^{\mu_1\dots \mu_{2n}}_{\nu_1\dots \nu_{2n}}R^{\nu_1\nu_2}_{\mu_1\mu_2}\dots R^{\nu_{2n-1}\nu_{2n}}_{\mu_{2n-1}\mu_{2n}}\, ,
%\end{equation}
 and the
$\lambda_n$ are arbitrary coupling constants with dimensions of length$^{2(n-1)}$. %and  we included, for completeness, a negative cosmological constant which is nonetheless irrelevant for the near-singularity behavior.

Static, spherically symmetric black holes in Lovelock theory take the form \req{bhfN} with $N(r)=1$
%\begin{eqnarray}\label{solQ}
%ds^2&=&-f(r)dt^2+\frac{dr^2}{f(r)}+r^2d\Sigma_{k,D-2}^2\, ,
%\end{eqnarray}
%where $d \Sigma^2_{k , D-2}$ is a constant curvature transverse geometry, with $k=+1, 0, -1$ denoting spherical, planar and hyperbolic, respectively.  T
and where the function $f(r)$ satisfies the algebraic equation
\begin{equation}\label{BHeq0}
h\left(\psi\right)=\frac{r_0^{D-3}}{r^{D-1}}\, , \qquad \text{where}\qquad \psi \equiv \frac{1-f(r)}{r^2}\, ,
\end{equation}
and where the ``characteristic polynomial'' $h(x)$ is given by \cite{Boulware:1985wk,Wheeler:1985nh,Myers:1988ze,deBoer:2009gx,Camanho:2011rj}
\begin{equation}
h(x )\equiv x+\sum_{n=2}^{\lfloor (D-1)/2\rfloor} \lambda_n x^n\, .
\end{equation}
%The integration constant $r_0$ is related to the mass $M$ via
%\begin{eqnarray}
%r_0^{D-3}&=&\frac{16\pi G M}{(D-2)\Omega_{D-2}}\,,
%\end{eqnarray}

%In this section we will explore the presence of Kasner eons in the interior of Lovelock black holes. Before doing so, let us start with some comments on the structure of black hole solutions for these theories.

%\subsection*{Four dimensions}
%The simplest non-trivial case corresponds to $D=4$. Then, the Lovelock Lagrangian simply reduces to the Einstein-Hilbert one, and the static and spherically symmetric solution is the Schwarzschild one,
%\begin{equation}
%f(r)=1-\frac{r_0}{r}\, ,
%\end{equation}
%which describes a black hole solution provided $r_0 > 0$, which we assume throughout the paper.
%In the deep interior, the effective Kasner exponents is given by
%\begin{equation}
%p_{\rm eff}(r)=-\frac{1}{3}\, .
%\end{equation}

%\subsection*{Five dimensions}

%\subsection*{Six dimensions}

%\subsection*{Seven dimensions}

Eq.\,\req{BHeq0} has $n$ solutions for $f(r)$. Of those, only one reduces in each case to the Schwarzschild one in the limit in which $\lambda_n\rightarrow 0$ $\forall$ $n$, and we will exclusively consider that one from now on. 
Now, the interior of a Lovelock black hole can be more complicated than in Einstein gravity. For example, even in the absence of charge or rotation it is possible to have an inner Cauchy horizon. Moreover, Lovelock black holes can have what are known as `branch singularities', which occur when a root of the polynomial $h(x)$ has a branch point~\cite{Kitaura:1990ve, Kitaura:1993cm}. While the metric remains finite, the radial derivatives of the metric function blow up at a branch singularity, meaning this is also a curvature singularity with divergent Kretschmann scalar. The volume of spatial slices does not become arbitrarily small at a branch singularity --- rather, it remains finite. A thorough analysis of the different types of interiors which arise as a function of the sign and magnitude of the gravitational couplings is presented in appendix \ref{GB}  for Gauss-Bonnet gravity in general dimensions as well as for cubic Lovelock gravity in $D=7$  --- namely, in the cases in which the Lovelock series is truncated at quadratic and cubic orders, respectively. %Here we do not analyze in detail the case of even higher-dimensional Lovelock theory. However, let us mention that 

%\comment{maybe some comments about qualitatively new stuff (such as three horizons) in $D\geq 8$}

Let us consider first the case in which the black holes contain a singularity at $r=0$.
%These exotic features are not relevant to the analysis we wish to perform, and so we restrict to coupling constants for which they do not occur. For example
This is generically the case if %$k = +1$ and 
${\rm sign}(\lambda_n) = +$ $\forall$ $n$. %\pbg{make sure}. 
%Additionally, in all cases we consider the solution which reduces to the $D$-dimensional Schwarzschild metric when ${\rm sign}(\lambda_n) \rightarrow 0$ $\forall$ $n$.
\begin{figure}
\centering \hspace{-.4cm}
\includegraphics[width=0.389\textwidth]{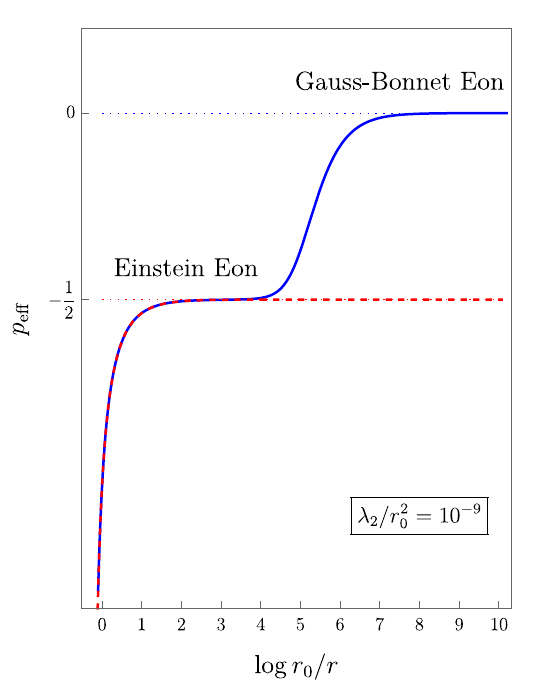}\hspace{-.5cm}
\includegraphics[width=0.339\textwidth]{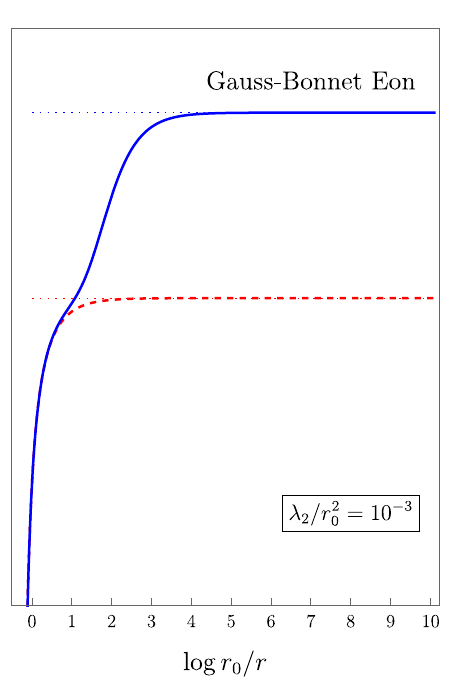}\hspace{-.5cm}
\includegraphics[width=0.339\textwidth]{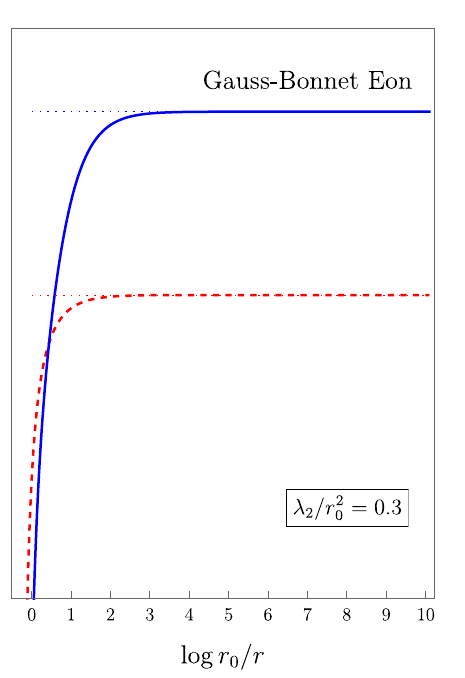}\\ \hspace{-.4cm}
\includegraphics[width=0.39\textwidth]{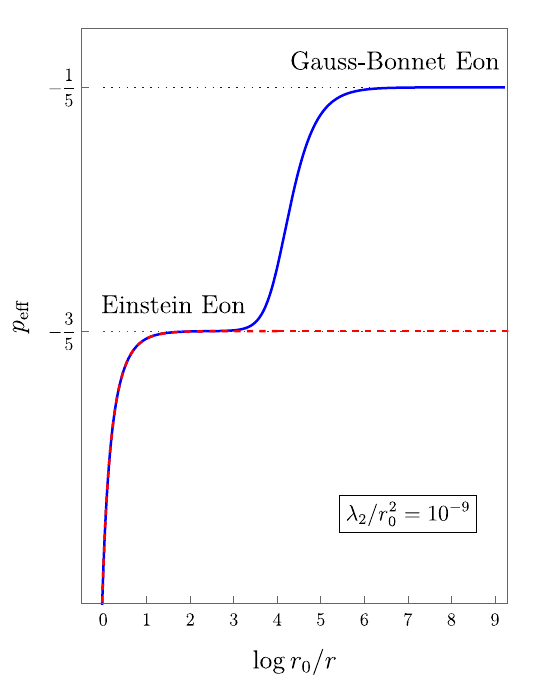}\hspace{-.5cm}
\includegraphics[width=0.339\textwidth]{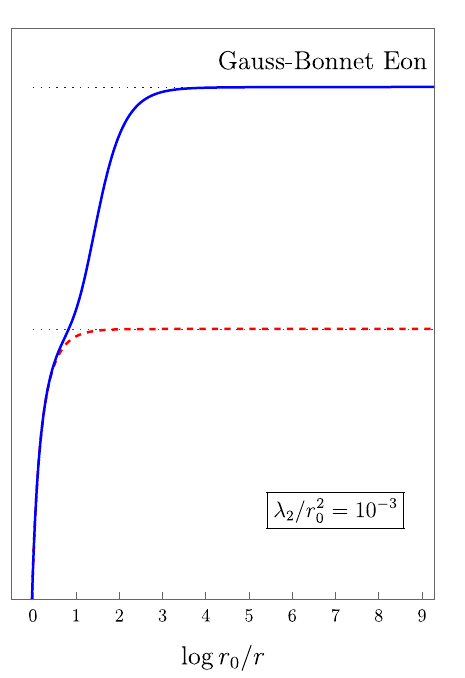}\hspace{-.5cm}
\includegraphics[width=0.339\textwidth]{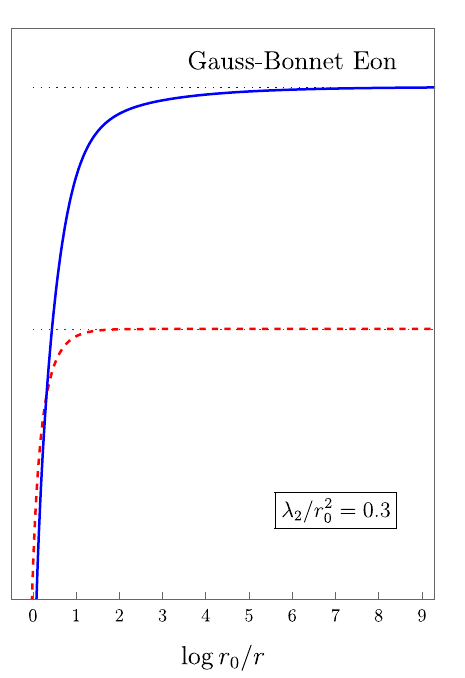}
\caption{We plot the effective Kasner exponent for Lovelock gravity black holes in $D=5$ (upper row) and $D=6$ (lower row) for various values of the Gauss-Bonnet coupling. For sufficiently small values of $\lambda_2/r_0^2$, the metric undergoes a Kasner eon characterized by the Einstein gravity exponent $p_{\rm eff}=-(D-3)/(D-1)$ and then transits to a new eon controlled by the Gauss-Bonnet density with $p_{\rm eff}=-(D-5)/(D-1)$ which characterizes the near-singularity metric. The dashed red curve corresponds to the usual $D$-dimensional Schwarzschild black hole.}
\label{fig:eon}
\end{figure}
\begin{figure}
\centering \hspace{-.4cm}
\includegraphics[width=0.389\textwidth]{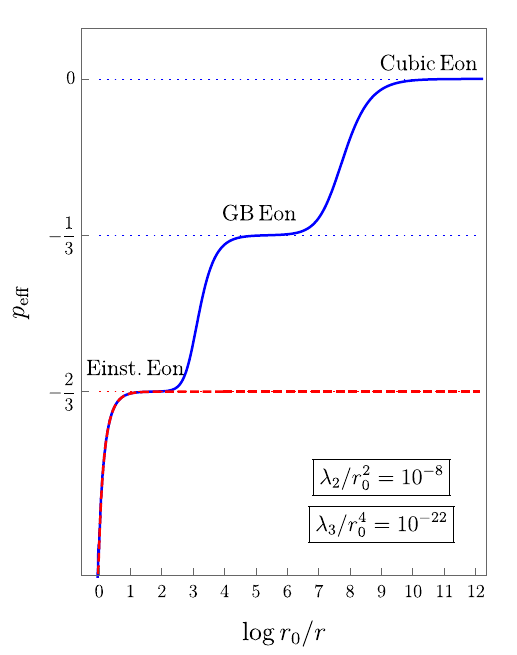}\hspace{-.5cm}
\includegraphics[width=0.339\textwidth]{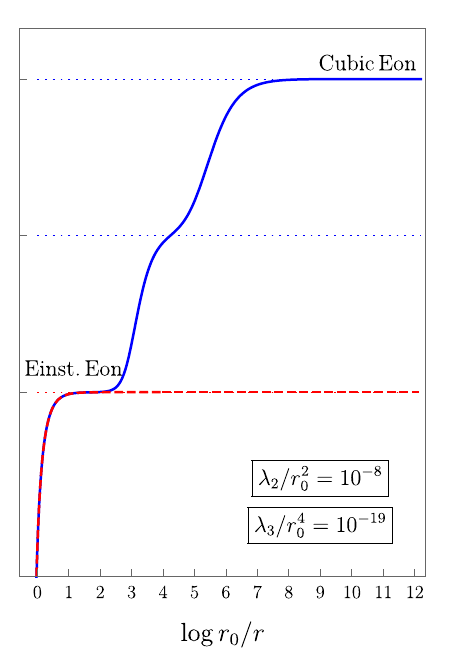}\hspace{-.5cm}
\includegraphics[width=0.339\textwidth]{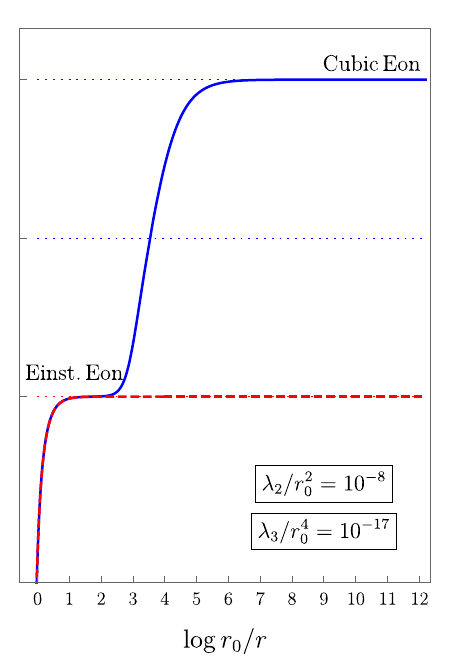}\\ 
\includegraphics[width=0.385\textwidth]{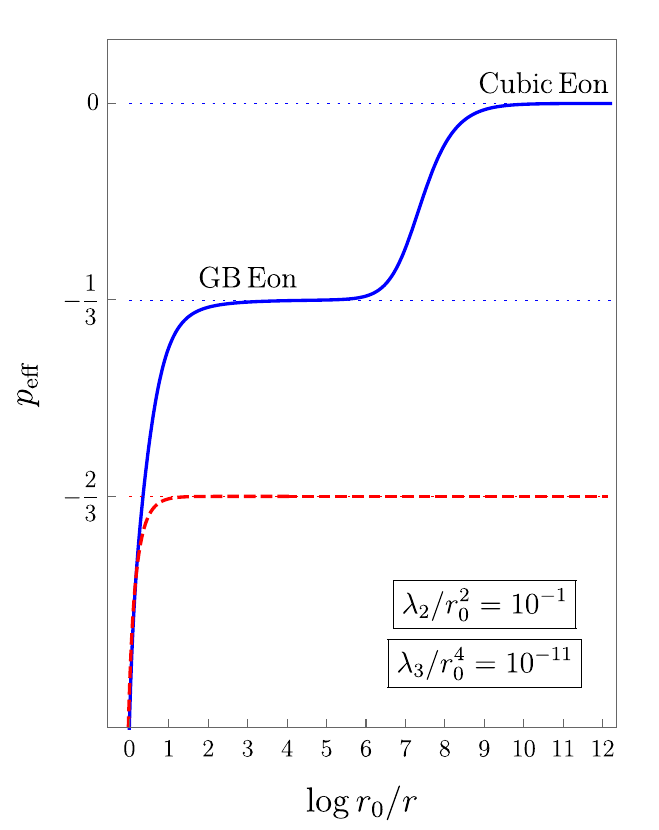}\hspace{-.6cm}
\includegraphics[width=0.338\textwidth]{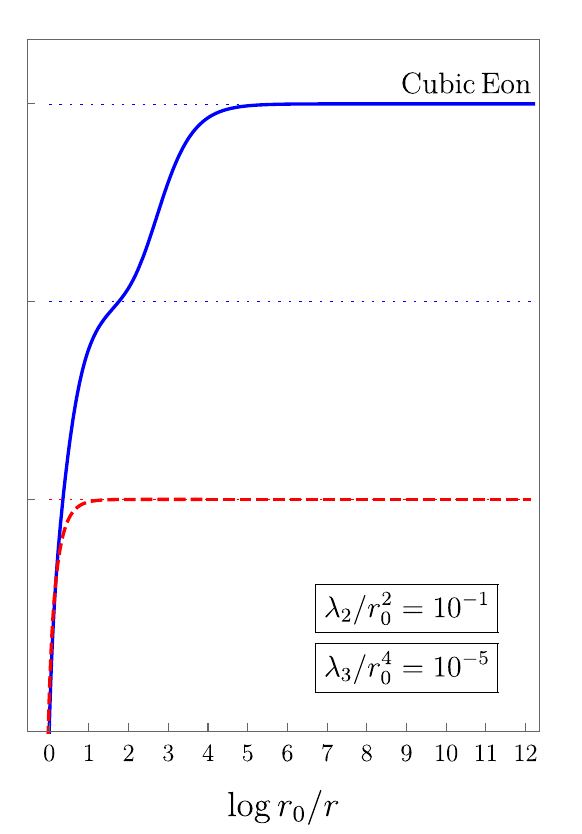}\hspace{-.6cm}
\includegraphics[width=0.338\textwidth]{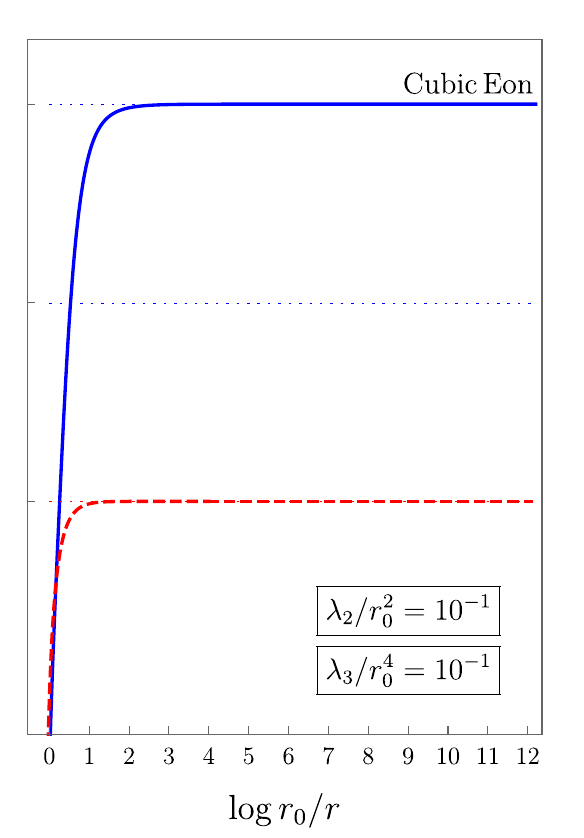}
\caption{We plot the effective Kasner exponent for Lovelock gravity black holes in $D=7$. Depending on the values of $|\lambda_3|/r_0^4$ and $|\lambda_2|/r_0^2$ it is possible to have three eons (upper left), an Einsteinian eon followed by a cubic Lovelock eon (upper right), a Gauss-Bonnet eon followed by a cubic Lovelock eon (lower left) or a single cubic Lovelock eon (lower right).  The dashed red curve corresponds to the $7$-dimensional Schwarzschild black hole.}
\label{fig:eon1}
\end{figure}
In the deep interior of such a black hole, it is only the highest-order density that contributes to the field equation. A simple computation shows that these Lovelock black holes have Kasner regimes in the deep interior which precisely correspond to the class of pure Kasner solutions of Lovelock gravity identified in \req{loveK} and \req{loveK2}.
%\begin{align}\label{LoveDeep}
%p_1 = - \frac{(D-2n-1)}{(D-1)} \, , \quad
%p_2 = \cdots = p_{D-1} = \frac{2n}{(D-1)} \, ,
%\end{align}
%which is precisely the class of pure Kasner solutions of Lovelock gravity identified in \req{loveK} and \req{loveK2}.
%\begin{align}
%\mu= 2n-1 \, ,
%\quad
%\nu = 1 + \frac{4 n(n-1)}{D-1} \, .
%\end{align}
In particular, note that those relations, combined with the fact that for Lovelock theory we must have $n \le (D-1)/2$, imply the following bounds on the Kasner exponents in general:
\be 
-1 \le p_1 \le 0 \, , \quad 0 \le p_{i \geq 2} \le 1 \, .
\ee
For $D=5$ and $D=6$, the Einsteinian eon --- characterized by approximately constant effective Kasner exponents with values given by \req{EinKas} --- is terminated when the Gauss-Bonnet term becomes dominant, and the effective Kasner exponents transition to approximately constant values given by  \req{loveK2} with $n=2$. If $\lambda_2$ is large enough, the Einsteinian eon can be completely skipped with $\peff$ transitioning directly to the Gauss-Bonnet phase. More precisely, if $\lambda_2$ is such that there exist a regime for which
\begin{equation}\label{ud1}
|\lambda_2|/r_0^2 \ll u^{D-1} \ll 1\, ,
\end{equation}
then there will be an Einsteinian eon corresponding to values of $r$ for which the above condition holds. On the other hand, if  $u^{D-1}$ becomes of the same order as $|\lambda_2|/r_0^2$ before $u^{D-1} \ll 1$ holds, the Einsteinian eon will be skipped and the transition will be directly to the Gauss-Bonnet one. These different cases are shown in particular examples for $D=5$ and $D=6$ in Fig.\,\ref{fig:eon}.

For $D\geq 7$, the last eon corresponds to  \req{loveK2} with $n=\lfloor (D-1)/2\rfloor$ and additional intermediate eons may arise (or be absent altogether) depending on the relative strength of the couplings. In case they arise, during each of those intermediate eons  \req{loveK2} holds with $n$ corresponding to the order of the density which is dominating the dynamics throughout that phase. The situation in which all possible intermediate eons arise requires that \req{ud1} holds for certain $u$, and that there exists a hierarchy of couplings of the form
%\begin{equation}
%|\lambda_n|/r_0^{2(n-1)} \ll \dots \ll |\lambda_3|/r_0^4 \ll |\lambda_2|/r_0^2\, .
%\end{equation}
\begin{equation}
|\lambda_n|^{1/(2(n-1))} \ll \dots \ll |\lambda_3|^{1/4} \ll |\lambda_2|^{1/2}\, .
\end{equation}
The various situations arising in the $D=7$ case are shown in Fig.\,\ref{fig:eon1}.

\begin{figure}
\centering \hspace{-.4cm}
\includegraphics[width=0.394\textwidth]{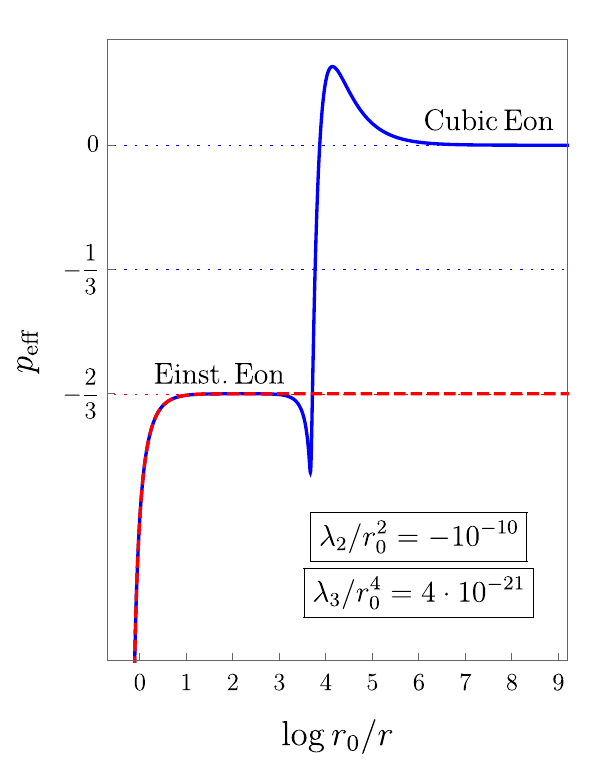}\hspace{-.5cm}
\includegraphics[width=0.336\textwidth]{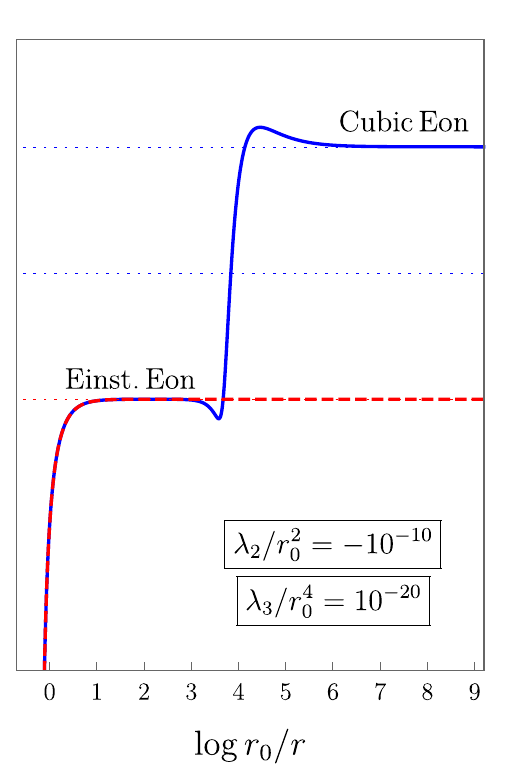}\hspace{-.5cm}
\includegraphics[width=0.336\textwidth]{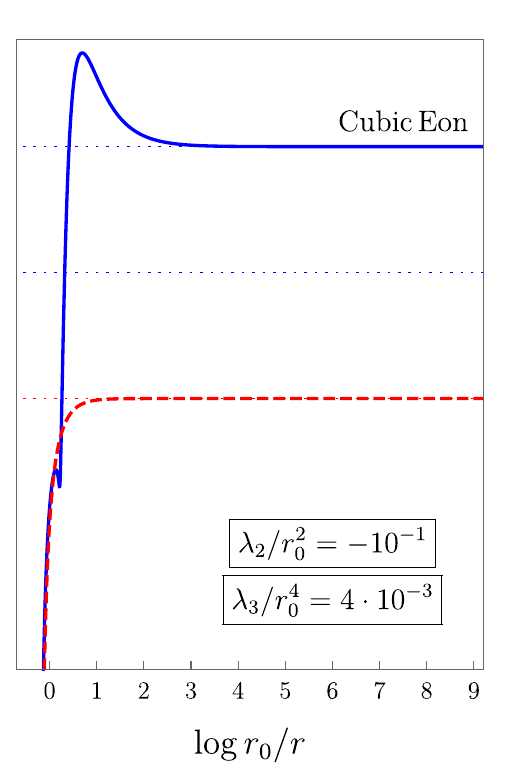}
\caption{We plot the effective Kasner exponent for Lovelock gravity black holes in $D=7$ for certain combinations of $\lambda_2$, $\lambda_3$ which give rise to a non-monotonic behavior of $p_{\rm eff}$. }
\label{fig:eon2}
\end{figure}

The cases considered so far are such that the effective Kasner exponent increases monotonically until it reaches a plateau corresponding to the final eon. However, for $D\geq 7$ it is possible to have transitions between eons which involve a non-monotonic behavior of $p_{\rm eff}$ and still conclude with a final eon which extends all the way to a singularity at $r=0$. For instance, this occurs in $D=7$ for $\lambda_2<0$, $1>\lambda_3/r_0^4>\lambda_2^2/(3r_0^4)$, as shown in some examples in Fig.\,\ref{fig:eon2}.

\begin{figure}
\centering \hspace{-.4cm}
\includegraphics[width=0.5\textwidth]{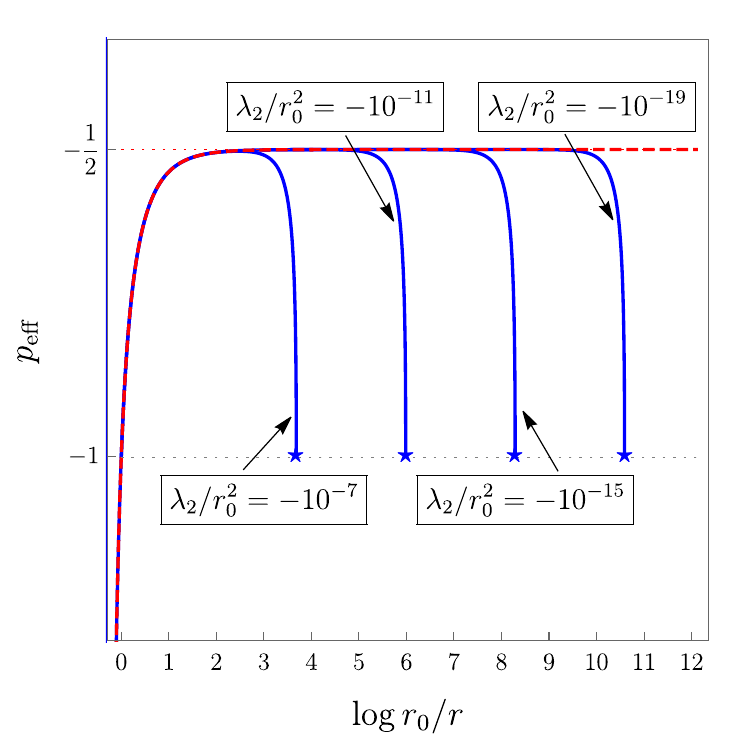}\hspace{-.5cm}
\includegraphics[width=0.5\textwidth]{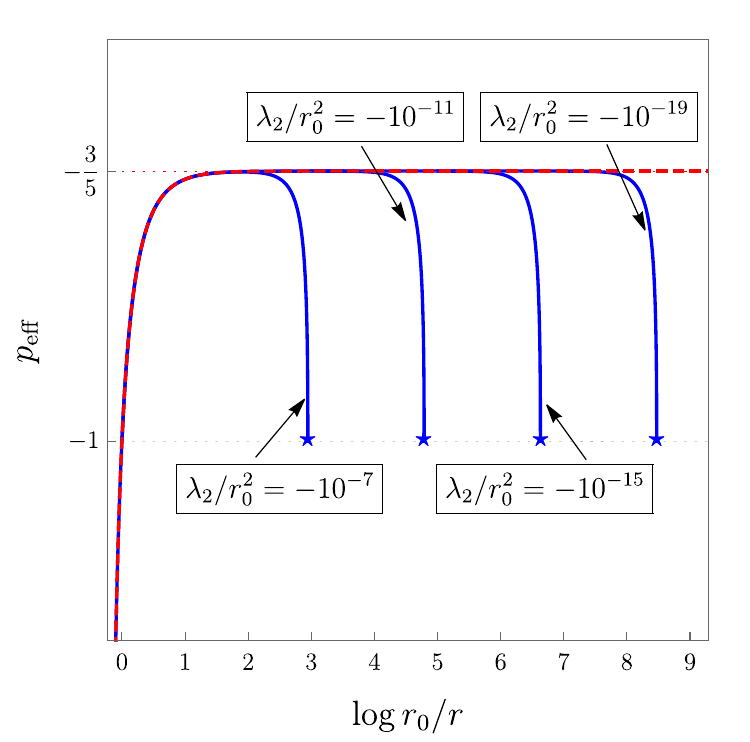}
\caption{We plot the effective Kasner exponent for Lovelock gravity black holes in $D=5$ (left) and $D=6$ (right) for various negative values of the Gauss-Bonnet coupling. For sufficiently small values of $|\lambda_2|/r_0^2$, an Einsteinian eon is present. Eventually, the effective Kasner exponent starts decreasing and takes the value $p_{\rm eff}=-1$ at the branch singularity $r=r_{\star}\equiv (4|\lambda_2|/r_0^2)^{1/(D-1)}$.
 }
\label{fig:eon3}
\end{figure}

We mentioned earlier that finite-volume singularities generically occur for Lovelock black holes for certain combinations of the couplings. In that case, an Einsteinian eon can still be present in the interior for sufficiently small values of the higher-curvature couplings. The radial derivative of the metric function $f(r)$ diverges at the finite-volume singularity, which we take to be at $r=r_{\star}$. Comparing with \req{effp}, it follows that
\begin{equation}
\lim_{r \rightarrow r_{\star}} f'(r) = \infty \quad \Rightarrow \quad p_{\rm eff}(r_{\star})=-1\, .
\end{equation}
Hence, in this case, the Einsteinian eon is followed by a decrease in $p_{\rm eff}$ which terminates at the singularity, where it takes the value $-1$ for general theories and dimensions. Examples of this behavior are shown in Fig.\,\ref{fig:eon3}.

% Let us study the transition between eons in more detail.
%\rah{Comments on eons}

%\begin{figure}
%\centering
%\includegraphics[width=0.65\textwidth]{eons_departure2.pdf}
%\caption{Kasner Eons for cubic Lovelock gravity in $D = 7$. The blue curve shows the effective Kasner exponent in the black hole interior. The light gray curves are the analytic results for Kasner exponents in Einstein, Gauss-Bonnet, and cubic Lovelock gravity in $D = 7$ (from bottom to top, respectively). The red lines give the analytic approximation for the end of an eon.}
%\label{fig:eon}
%\end{figure}

\section{The end of an eon} 
\label{end_eon}
In this section we have a closer look at how a Kasner eon comes to an end. We provide a perturbative formula for the effective Kasner exponent at the end of a Lovelock gravity eon and make some general comments about the termination of the Einsteinian one. In addition, we show that if the effective stress tensor generated by higher-curvature terms satisfies the null energy condition, then the effective Kasner exponent exhibits a monotonic behaviour at the end of the Einsteinian eon.

%Some general comments can be made regarding how a Kasner eon ends
\subsection{The end of an eon in Lovelock theory}

As we have seen, during an eon the effective Kasner exponent is approximately constant. However, if higher-curvature terms are present then eventually these will become important and drive the universe to a new eon, as shown in the figures above. In Lovelock theory, we can derive analytically the leading corrections to $\peff$ in that regime. 

Since it is easy to do so, let us consider the following situation. The universe is in an eon where the Lovelock term of order $n$ dominates. We consider a transition between the order $n$ eon and an order $m$ eon. The coupling $\lambda_n$ is treated non-perturbatively, while we compute only the leading correction for $\lambda_m$. The result of this computation is
\begin{align} 
\peff =&\, - \frac{D-2n-1}{D-1} + \frac{2 (m-n)}{D-1}  \frac{|\lambda_{m}| u^{(D-1) \left(1- m/n\right)}}{|\lambda_{n}|^{m/n} r_0^{2(m/n-1)}} + \cdots
\end{align}
An eon can be considered to have ended when the second term in the above becomes $\mathcal{O}(1)$. This will happen at the point where
\be 
 \frac{|\lambda_{m}| \, u^{(D-1) \left(1- m/n\right)}}{|\lambda_{n}|^{m/n} \, r_0^{2(m/n-1)}}\sim 1 \, .
\ee
Naturally,  the most interesting case is the end of the Einstein gravity eon (this is also the case that is within reach of conventional effective field theory). So we consider the case where $n = 1$ and allow $m$ to remain arbitrary. It is convenient to introduce a scale $\ell$ for the coupling $\lambda_m$ so that $\lambda_m \sim \mu_m \ell^{2m-2}$ where $\mu_m$ is dimensionless and order one and $\ell$ is a length scale. The result is that the Einstein eon ends at
\be 
r_{\rm end} \sim r_0 \left(\frac{\ell}{r_0}\right)^{2/(D-1)} \, ,
\ee
which is independent of $m$. That is, the result is the same if it is Gauss-Bonnet gravity taking over, or the twelfth-order Lovelock taking over. Notably, because of the fractional exponent $2/(D-1)$ this point can be \textit{orders of magnitude} larger than the length scale $\ell$ that characterizes the new physics.\footnote{This feature has been emphasized in~\cite{Martinec:1994xj, Zigdon:2024ljl}.} The result is more intuitive when expressed in terms of the proper time. Since the proper time during the Einstein gravity eon is 
\be 
\tau \sim r_0 u^{(D-1)/2} \, ,
\ee
we have $\tau_{\rm end} \sim \ell$. %\rah{Another way to see that same thing is to take $R_{abcd}^2$ for Schwarzschild and set, e.g., $R_{abcd}^2 \sim r_0^2/r^6 \sim 1/\ell^4$.}

\subsection{Monotonicity of the effective Kasner exponent}

%As we saw before, $\peff $ behaves in a monotonically-increasing fashion in most (but not all) situations in which the Schwarzschild singularity does not get replaced by a finite-volume one. On the other hand, the presence of a finite-volume singularity enforces the condition $p_{\rm eff}'(r)>0$ at the end of the Einsteinian eon\footnote{Note that in the figures above moving to the right means moving towards decreasing values of $r$}. \comment{Perhaps this could be used to test the possible fate of interiors in perturbative scenarios by looking at the sign of $p_{\rm eff}'(r)$ right at the end of the Einsteinian eon.} \comment{I had removed Robie's lines with the approximation to the end of an eon, perhaps put back/make additional plots}

In the examples we have studied we have seen that the effective Kasner exponent often --- but not always --- \textit{increases} as one moves toward the singularity. Moreover, we saw that the finite volume singularities were always associated with an effective Kasner exponent that \textit{decreases} at the end of an eon. Here we will put these observations onto a somewhat more rigorous footing, making a connection between the monotonicity of the effective Kasner exponent and the null energy condition. 

The null energy condition requires that $T_{\mu\nu} k^\mu k^\nu \ge 0$ for all null vectors $k$. Here we are considering vacuum spacetimes of higher-curvature theories. While, strictly speaking, there is no matter in the setup, we can consider the higher-curvature terms to generate an \textit{effective} stress-energy tensor $T_{\mu\nu} = G_{\mu\nu}$.  For a static and spherically symmetric spacetime characterized by a single metric function $f(r)$ the null energy condition implies the following constraint:
\be 
G_{\mu\nu} k^\mu k^\nu \ge 0 \quad \forall k^\mu \quad \Rightarrow \quad r^2 f'' + (D-4) r f' + 2(D-3)(1-f) \ge 0 \, .
\ee

Let us now study the end of the Einsteinian eon. Consider a correction of the form
\be 
f(r) = 1 - \left(\frac{r_0}{r} \right)^{D-3} + \lambda \left(\frac{r_0}{r} \right)^s \, ,
\ee
where $\lambda$ is a coupling parameter and $s > D-3$ so that the correction is subleading to the Einstein terms. We are considering this not as an exact solution but instead as a model of the leading (in $\lambda$) correction to the Schwarzschild solution. Plugging this into the constraint above, we find
\be\label{nec_eval} 
r^2 f'' + (D-4) r f' + 2(D-3)(1-f) = \lambda (s-2)(s-D+3) \left(\frac{r_0}{r} \right)^s \, ,
\ee
indicating that the null energy condition is satisfied provided that $\lambda > 0$. 

Next consider the derivative of the effective Kasner exponent at the end of the Einsteinian eon. Expanding to leading order in $\lambda$ and working in the $r \ll r_0$ limit, we find
\be 
p'_{\rm eff}(r) = - \frac{2 \lambda (s - D + 3)^2  }{r_0 (D-1)^2}  \left(\frac{r_0}{r} \right)^{D - 4 - s}\, .
\ee
To make the result more transparent and consistent with the plots, let us introduce the coordinate $y \equiv \log(r_0/r)$ which \textit{increases} toward the singularity. In terms of this coordinate, we obtain
\be 
\frac{d p_{\rm eff}}{dy} = \frac{2  \lambda (s + D - 3)^2}{(D-1)^2} e^{(s - D + 3) y} \, .
\ee
This means that if the null energy condition is satisfied, then the effective Kasner exponent must increase at the end of the Einsteinian eon. On the other hand, if the effective Kasner exponent is seen to \textit{decrease} toward the singularity, then this indicates a violation of the null energy condition. This latter result was seen to be universally associated with the finite volume singularities of Gauss-Bonnet gravity studied in the earlier sections --- see Figure~\ref{fig:eon3}.

The effective Kasner exponent is negative in Einstein gravity and governs the expansion of spacetime along the $z$ direction in the black hole interior. The above result tells us that if the corrections to General Relativity respect the null energy condition then this expansion is \textit{slowed} as a result of the corrections. 

A natural question is whether the null energy condition can tell us anything about the monotonicity of the $p_{\rm eff}$ at the end of other Lovelock eons beyond the Einsteinian one. Unfortunately, the answer appears to be no. It would be interesting to assess whether other constraints can yield useful insights in this case.

\section{Discussion}

Motivated by the key role they play in the approach to a spacelike singularity, we began our work by classifying the types of Kasner solutions that can arise in higher-curvature theories of gravity. In general, the conditions on the Kasner exponents differ significantly from Einstein gravity. We noted a universal feature of Kasner metrics in higher-curvature gravity: For a theory incorporating $n$ powers of curvature, there always exists a Kasner solution for which the exponents satisfy
\be \label{pix}
\sum_{i=1}^{D-1} p_i = 2 n  - 1 \, .
\ee
In the case of Einstein gravity $n = 1$ and the well-known condition on the sum of the Kasner exponents is recovered. For General Relativity, the above is the unique condition on the sum of the Kasner exponents dictated by the field equations. However, in higher-curvature theories there can be additional families of solutions beyond this universal one. It is nonetheless natural to speculate that the Einstein gravity condition, $\sum_{i=1}^{D-1} p_i (x)= 1$, satisfied at each spatial point in the approach to a generic singularity, would be replaced by \req{pix} with spatially dependent exponents in the case of Kasner eons dominated by order-$n$ densities. This would rely on the persistence of ultralocality in those cases, a feature which remains to be explored.

We have further explored the concept of a Kasner eon first introduced in~\cite{Bueno:2024fzg}, focusing on the example of Lovelock gravity. We have provided a detailed analysis of the interior structure of Lovelock black holes and analysed examples where additional eons or finite volume singularities occur in the interior. A key result of our analysis concerns the monotonicity of the effective Kasner exponent at the end of the Einsteinian eon. We demonstrated that if the effective stress tensor generated by the higher-curvature corrections obeys the null energy condition, then the effective Kasner exponent must increase at the end of the Einsteinian eon. Since this exponent (when negative) controls the expanding direction of the universe, the physical implication is that the null energy condition demands that the expansion in this direction \textit{slows} as the singularity is approached. The net result is that, very near the singularity, the spatial volume of the universe collapses more slowly, scaling like
\be 
V \sim \tau^{1 + \delta  (D-1)}  \, , \quad p_{\rm eff} = - \frac{(D-3)}{(D-1)} + \delta \, .
\ee

Going forward it will be important to understand the holographic implications of Kasner eons. Here we have focused on the asymptotically flat setting, but our results will carry over to the asymptotically AdS case as well. This is because in the deep black hole interior the negative cosmological constant becomes irrelevant. As one example, the existence of eons can explain the confusion that originally arose concerning applications of the ``Complexity = Action'' proposal~\cite{Brown:2015bva} to higher-curvature black holes. It was observed that even when the higher-curvature couplings are turned off, the late-time growth rate of complexity does not reduce to its Einstein gravity value~\cite{Cai:2016xho, Cano:2018aqi, Emparan:2021hyr}. Ultimately, this is because the late-time growth rate of complexity is sensitive to the final eon in the black hole interior. In fact, the late-time complexity growth rate in Lovelock theory (in $D > 2 n + 1$) can be expressed in terms of the effective Kasner exponent~\cite{CHM},
\be 
\lim_{t\to\infty} \frac{d \mathcal{C}}{dt} = \frac{(D-1) (1 + p_{\rm eff}) M}{\pi \left(D-2 + (D-1)p_{\rm eff} \right)} \, .
\ee
Therefore, because the exponents governing the final eon in Lovelock theory are different from Einstein gravity, the growth rate is different. And because the Kasner exponents are always constants independent of the couplings, the limit of the growth rate does not recover the Einstein gravity result. Substituting $p_{\rm eff} = -(D-3)/(D-1)$ into the above, one recovers the well-known $2 M/\pi$ predicted by General Relativity. Further noting that the null energy condition requires that $p_{\rm eff}$ should \textit{increase} at the end of the Einsteinian eon, one concludes that the complexification rate should \textit{decrease} as additional eons are probed. It would be particularly interesting to revisit this analysis, considering for example the time dependence of complexity which may exhibit distinct features as the Wheeler-DeWitt patch picks up contributions from different eons.

The interiors of spherically symmetric black holes provide a simple consistency check of the concept of eons. This is because one often has access to the solution \textit{exactly}. However, it will be important to test the concept of an eon under less symmetric conditions. Ultimately, the idea is that one may have additional BKL-like phases of evolution driven by higher-curvature or quantum corrections to the Einstein equations. It's therefore essential to explore these ideas as generically as possible without becoming too reliant on highly symmetric examples. We hope to return to this problem in the near future.
%From this perspective, the most approachable and natural question concerns the end of the Einsteinian eon which can be addressed using conventional EFT methods. 
\\
\\
\\
\noindent
{\bf Note Added}: When we were in the final stages of preparing this manuscript~\cite{Caceres:2024edr} appeared on the arXiv. That paper develops the ideas of Kasner eons in a manner complementary to our own, by exploring the role of matter and focusing on the case of quasi-topological gravities. Those authors also perform a preliminary investigation of the holographic interpretation of Kasner eons.

\section*{Acknowledgments}

We are grateful to {\'A}ngel Murcia and Yoav Zigdon for helpful comments and discussions. 
PB was supported by a Ram\'on y Cajal fellowship (RYC2020-028756-I) from Spain's Ministry of Science and Innovation and by
the  grant PID2022-136224NB-C22, funded by MCIN/AEI/ 10.13039/501100011033/FEDER, UE. The work of PAC received the support of a fellowship from “la Caixa” Foundation (ID 100010434) with code LCF/BQ/PI23/11970032.
The work of RAH received the support of a fellowship from ``la Caixa” Foundation (ID 100010434) and from the European Union’s Horizon 2020 research and innovation programme under the Marie Skłodowska-Curie grant agreement No 847648 under fellowship code LCF/BQ/PI21 /11830027. MDL was supported by a Chinese Scholarship Council (CSC) fellowship.

\appendix 

\section{The interior of Lovelock black holes}\label{GB}
\subsection{Gauss-Bonnet in general $D$}
Truncating the Lovelock action at quadratic order yields the Einstein-Gauss-Bonnet action
\begin{equation}\label{LoveF}
I=\frac{1}{16\pi G}\int_{\mathcal{M}}\mathrm{d}^{D}x\sqrt{|g|} \left[ R + \frac{\lambda_2}{(D-3)(D-4)} \left(R^2- 4R_{ab}R^{ab}+R_{abcd}R^{abcd}\right) \right]\, ,
\end{equation}
where again we set the cosmological constant to zero and where $\lambda_2$ has dimensions of length$^2$. The Gauss-Bonnet term is dynamical for $D\geq 5$, topological in $D=4$ and trivially zero for $D\leq 3$. 

In $D\geq 5$, the theory admits static and spherically symmetric black hole solutions characterized by a single metric function which satisfies
\begin{equation}\label{BHeq}
h\left(\psi\right)=\frac{r_0^{D-3}}{r^{D-1}}\, , \qquad \text{where}\qquad \psi \equiv \frac{1-f(r)}{r^2}\, ,
\end{equation}
where we always assume $r_0>0$, and where the ``characteristic polynomial'' $h(x)$ is given by 
\begin{equation}
h(x)\equiv x + \lambda_2 x^2\, .
\end{equation}
This equation has two solutions. We consider the one which has a well-defined Einstein-gravity limit when $\lambda_2/r_0^2 \rightarrow 0$. This reads
\begin{equation}
f(r)=1+\frac{r^2}{2\lambda_2}\left[1 - \sqrt{1+\frac{4 \lambda_2 r_0^{D-3}}{r^{D-1}}} \right] \overset{(\lambda_2/r_0^2\rightarrow 0)}{=} 1 - \left(\frac{r_0}{r}\right)^{D-3}+\left(\frac{r_0}{r}\right)^{2(D-2)} \left(\frac{\lambda_2}{r_0^2}\right)+ \dots
\end{equation}
Whenever  
\begin{align}
&0\leq \left(\frac{\lambda_2}{r_0^2}\right) <1 && (D=5)\, , \\
 &0 \leq \left(\frac{\lambda_2}{r_0^2}\right) && (D\geq 6)\, ,
\end{align}
 this function has a single real zero, $f(r_{\rm h})=0$, $r_{\rm h}>0$. In all such cases, the solution describes a black hole with a curvature singularity at $r=0$ hidden behind an event horizon at $r=r_{\rm h}$.  %\comment{The singularity is softened with respect to the Schwarzschild one. Write the dependence in powers of $r$ of the Kretschmann invariant for general $D$} 
 The explicit form of $r_{\rm h}$ for the first few dimensions reads
\begin{align}
r_{\rm h}&=\sqrt{1-\left(\frac{\lambda_2}{r_0^2}\right)} &&(D=5) \, , \\
r_{\rm h}&=\frac{\left[9+\sqrt{3}\sqrt{27+4\left(\frac{\lambda_2}{r_0^2}\right)^3}\right]^{1/3}}{2^{1/3}3^{2/3}}-\frac{2^{1/3} \left(\frac{\lambda_2}{r_0^2}\right)}{3^{1/3} \left[9+\sqrt{3}\sqrt{27+4\left(\frac{\lambda_2}{r_0^2}\right)^3}\right]^{1/3}} &&(D=6)\, , \\
r_{\rm h}&=\frac{1}{\sqrt{2}} \sqrt{-\left(\frac{\lambda_2}{r_0^2}\right)+\sqrt{4+\left(\frac{\lambda_2}{r_0^2}\right)^2} } &&(D=7)\, .
\end{align}
On the other hand, in general $D\geq 5$,  whenever 
\begin{equation}
-\frac{1}{2^{\frac{D-5}{D-3}}}< \left(\frac{\lambda_2}{r_0^2}\right) <0\, ,
\end{equation}
the solution describes a black hole hidden behind an event horizon at $r=r_{\rm h}$ which has a finite-volume singularity at
\begin{equation}
r_{\star}\equiv \left(\frac{4 |\lambda_2|}{r_0^2}\right)^{\frac{1}{D-1}}\, .
\end{equation}
Finally, whenever 
\begin{equation}
 \left(\frac{\lambda_2}{r_0^2}\right) \leq  -\frac{1}{2^{\frac{D-5}{D-3}}}\, ,
\end{equation}
the solution describes a naked singularity at $r_{\star}$. Additionally, in the special case of $D=5$, a naked singularity at $r=0$ also arises for 
\begin{equation}
1 \leq  \left(\frac{\lambda_2}{r_0^2}\right) \, .
\end{equation}

\subsection{Cubic Lovelock in $D=7$}
\begin{figure}
\centering \hspace{-.4cm}
\includegraphics[width=0.6\textwidth]{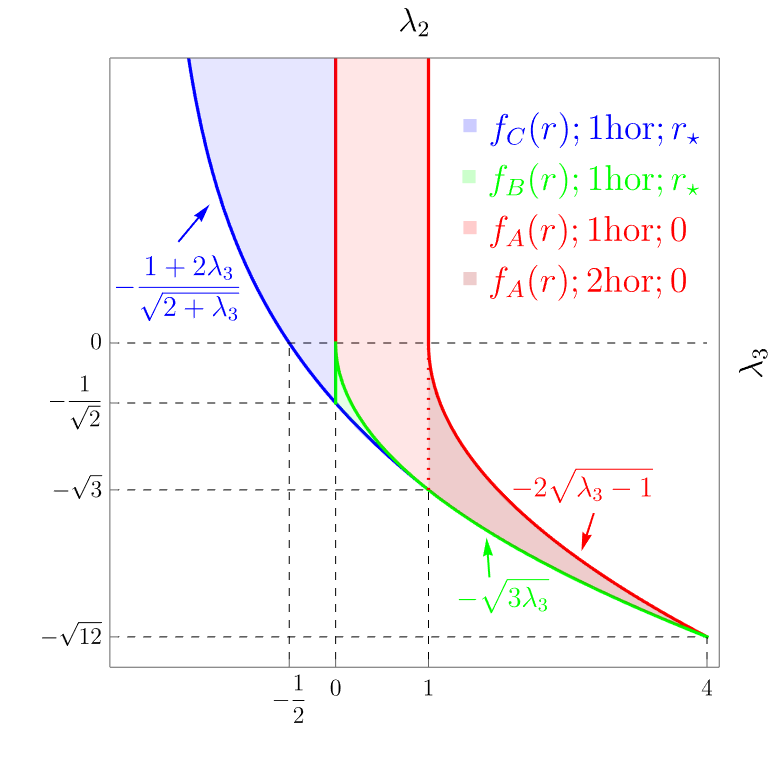}
\caption{We plot the space of black hole solutions of  the cubic Lovelock gravity with Lagrangian \req{LoveF2} parametrized by the values of $\lambda_2/r_0^2$ and $\lambda_3/r_0^4$ (we omit $r_0$ everywhere to avoid the clutter). The blue region corresponds to black holes described by $f_C(r)$ and corresponds to black holes with a single horizon and with a finite-volume singularity at $r_\star$. The green region corresponds to black holes whose metric function is $f_B(r)$ and which possess a single horizon and a finite-volume singularity at $r_\star$. The lighter red region corresponds to black holes with a metric function given by $f_A(r)$ and which possess a single horizon and a singularity at $r=0$. Finally, the darker red region also contains black holes described by $f_A(r)$ with a singularity at $r=0$ but with two horizons.
 }
\label{figo3}
\end{figure}

Consider now the case of a cubic Lovelock theory in $D=7$. The Lagrangian reads
\begin{equation}\label{LoveF2}
I=\frac{1}{16\pi G}\int_{\mathcal{M}}\mathrm{d}^{7}x\sqrt{|g|} \left[ R + \frac{\lambda_2}{12}\mathcal{X}_4+ \frac{\lambda_3}{24} \mathcal{X}_6\right]\, ,
\end{equation}
where $\lambda_2$ and $\lambda_3$ have dimensions of length$^2$ and length$^4$ respectively. The theory admits static and spherically symmetric black hole solutions characterized by a single metric function which now satisfies
\begin{equation}\label{BHeq}
h\left(\psi\right)=\frac{r_0^{4}}{r^{6}}\, , \qquad \text{where}\qquad \psi \equiv \frac{1-f(r)}{r^2}\, ,
\end{equation}
and where the ``characteristic polynomial'' $h(x)$ is now given by 
\begin{equation}
h(x)\equiv x + \lambda_2 x^2+ \lambda_3 x^3\, .
\end{equation}
This equation has three solutions which can be written as
\begin{align}
f_A&\equiv \frac{1}{3\lambda_3} \left[(3\lambda_3+\lambda_2 r^2)+\frac{2^{1/3}r^4(3\lambda_3-\lambda_2^2)}{(\Sigma+3\sqrt{3}\sqrt{\Upsilon})^{1/3}} -\frac{(\Sigma+3\sqrt{3}\sqrt{\Upsilon})^{1/3}}{2^{1/3}}\right] \, , \\
f_B&\equiv \frac{1}{3\lambda_3} \left[(3\lambda_3+\lambda_2 r^2)-\frac{(1+ {\rm i} \sqrt{3})r^4(3\lambda_3-\lambda_2^2)}{2^{2/3}(\Sigma+3\sqrt{3}\sqrt{\Upsilon})^{1/3}} +\frac{(1-{\rm i} \sqrt{3})(\Sigma+3\sqrt{3}\sqrt{\Upsilon})^{1/3}}{2^{4/3}}\right]   \, , \\
f_C&\equiv \frac{1}{3\lambda_3} \left[(3\lambda_3+\lambda_2 r^2)-\frac{(1-{\rm i} \sqrt{3})r^4(3\lambda_3-\lambda_2^2)}{2^{2/3}(\Sigma+3\sqrt{3}\sqrt{\Upsilon})^{1/3}} +\frac{(1+{\rm i} \sqrt{3})(\Sigma+3\sqrt{3}\sqrt{\Upsilon})^{1/3}}{2^{4/3}}\right]  \, ,
\end{align}
where
\begin{equation}
\Upsilon(r)\equiv 27\lambda_3^4-4\lambda_2^3\lambda_3^2 r^6+18 \lambda_2 \lambda_3^3 r^6-\lambda_2^2\lambda_3^2 r^{12}+4\lambda_3^3 r^{12}\, , \,\,\, \Sigma(r)\equiv 27\lambda_3^2-2\lambda_2^3 r^6+ 9 \lambda_2 \lambda_3 r^6\, ,
\end{equation}
and where we set $r_0=1$ in all expressions\footnote{This can be easily reintroduced by replacing $\lambda_2\rightarrow \lambda_2/r_0^2$ and $\lambda_3\rightarrow \lambda_3/r_0^4$.}. The region in parameter space for which asymptotically flat black holes exist is displayed in Fig.\,\ref{figo3}. 
Whenever 
\begin{equation}
-\frac{1+2\lambda_3}{\sqrt{2+\lambda_3}}<\lambda_2\quad \text{and }\quad \lambda_3\leq 1
\end{equation}
there exists a black hole with a single horizon at
\begin{equation}
r_{\rm h}=\sqrt{\frac{-\lambda_2+\sqrt{4+\lambda_2^2-4\lambda_3}}{2}}\, .
\end{equation}
In particular, when 
\begin{equation}
-\frac{1+2\lambda_3}{\sqrt{2+\lambda_3}}<\lambda_2\quad \text{and }\quad \lambda_3< 0
\end{equation}
the solution is described by $f_C(r)$ above and there is a finite-volume singularity at
\begin{equation}
r_\star \equiv \left( \frac{-2\lambda_2^3+2(\lambda_2^2 -3\lambda_3)^{3/2}+9\lambda_2\lambda_3}{\lambda_2^2-4\lambda_3} \right)^{1/6}\, .
\end{equation}
Also, when 
\begin{equation}
-\frac{1+2\lambda_3}{\sqrt{2+\lambda_3}}<\lambda_2<-\sqrt{3\lambda_3} \quad \text{and}\quad 0< \lambda_3\, ,
\end{equation}
the solution is described by $f_B(r)$ and there is a finite-volume singularity at $r_\star$. Finally, when 
\begin{equation}
0\leq \lambda_3 <1 \quad \text{and} \quad -\sqrt{3\lambda_3}<\lambda2
\end{equation}
the solution is described by $f_A(r)$ and it contains a singularity at $r=0$.

When
\begin{equation}
1<\lambda_3<4\quad \text{and}\quad -\sqrt{3\lambda_3}<\lambda_2< -2\sqrt{\lambda_3-1}\, ,
\end{equation}
the solution is described by $f_A(r)$, it contains a Cauchy horizon and an event horizon, respectively at
\begin{equation}
r_{h_{\rm C}}= \sqrt{\frac{-\lambda_2-\sqrt{4+\lambda_2^2-4\lambda_3}}{2}} \, , \quad r_{h}= \sqrt{\frac{-\lambda_2+\sqrt{4+\lambda_2^2-4\lambda_3}}{2}}\, ,
\end{equation}
and a singularity at $r=0$.

\bibliographystyle{JHEP-2}
\bibliography{kasner.bib}

\end{document}